\documentclass[letterpaper,10pt,conference]{IEEEtran}
\IEEEoverridecommandlockouts                 

\usepackage{graphics} 
\usepackage{epsfig} 
\usepackage{times} 
\usepackage{amsmath} 
\usepackage{upgreek} 
\newcommand{\norm}[1]{\| #1 \|}
\usepackage{amsfonts}
\usepackage{amssymb}  
\usepackage{bm}

\usepackage{enumitem}

\usepackage[utf8]{inputenc}
\usepackage[english]{babel}
\usepackage[dvipsnames]{xcolor}
\usepackage[version-1-compatibility]{siunitx}

\usepackage{amsmath}               
  {
    \newtheorem{theorem}{Theorem}
    
    \newtheorem{assumption}{Assumption}
    \newtheorem{remark}{Remark}
  }

\makeatletter
\newcommand*\bigcdot{\mathpalette\bigcdot@{1}}
\newcommand*\bigcdot@[2]{\mathbin{\vcenter{\hbox{\scalebox{#2}{$\m@th#1\bullet$}}}}}
\makeatother

\begin{document}
\newcommand{\cmmnt}[1]{}
\newcommand\blfootnote[1]{%
  \begingroup
  \renewcommand\thefootnote{}\footnote{#1}%
  \addtocounter{footnote}{-1}%
  \endgroup
}

\title{\LARGE \bf Leveraging PID Gain Selection Towards Adaptive Backstepping Control for a Class of Second-Order Systems}
%

\author{Ahmad~Kourani$^{1}$
        and Naseem Daher$^{2}$, ~\IEEEmembership{Member,~IEEE}
\thanks{$^{1}$Ahmad Kourani is with the Vision and Robotics Lab, Department of Mechanical Engineering, 
American University of Beirut, Beirut, Lebanon
        {\tt\small ahk42@mail.aub.edu}}%

\thanks{$^{2}$Naseem Daher is with the Vision and Robotics Lab, Department of Electrical \& Computer Engineering,
       {\tt\small nd38@aub.edu.lb}}%
       }

\IEEEoverridecommandlockouts

\IEEEpubid{\makebox[\columnwidth]{\copyright{}2021 IEEE. Personal use of this material is permitted. \hfill} \hspace{\columnsep}\makebox[\columnwidth]{ }}

\maketitle

\begin{abstract}
In this work, we establish a convenient similarity between an adaptive backstepping control law and a standard proportional-integral-derivative (PID) controller for a class of second-order systems. The extracted similarity provides a deeper understanding of the adaptive backstepping design from a performance perspective via an intuitive method to select its otherwise abstract controller gains, on top of its traditional stability perspective. Such a similarity analysis opens the door for researchers to use well-established PID tuning methods to predict the performance of Lyapunov stability-based controllers. At the same time, the obtained formulation reveals how the corresponding PID control law can be linked to Lyapunov stability theory. 
The proposed scheme is applied to a quadrotor unmanned aerial vehicle (UAV) carrying a payload
with the presence of a wind gust as an external disturbance, in simulation and experimentally.
\end{abstract}
\begin{IEEEkeywords}
Adaptive Backstepping Control, Robust Control, Lyapunov Stability, PID, 
Quadrotor, Payload Variation.
\end{IEEEkeywords}
%
\section{Introduction} \label{introduction}
Proportional-Integral-Derivative (PID) control systems dominate real-world industrial applications owing to their effectiveness and simple architecture,  which entails determining only three gains \cite{Chang2002SelfTuningPID}.
Furthermore, practical tuning methods exist to tune the PID gains, such as the widely-used Ziegler–Nichols tuning method \cite{Astrom1995}, in addition to other newly established equivalence-based gain selection and tuning techniques \cite{Chang2009_PID_gain_selection,Lee2014_PID_Backstepping_tuning}.
However, suffering from oversimplification amongst other issues \cite{Han2009PIDtoADRC}, PID techniques continued to evolve to deal with more complex systems including nonlinear and time-varying ones.
\par
More advanced designs that outperform PID controllers exist in the literature, however, they are more complex and require deeper theoretical knowledge, which presents a challenge towards their wide adoption.
Backstepping control is a common and widely-used approach to deal with nonlinear systems. 
While conventional backstepping and modified backstepping achieve asymptotic stability \cite{Yu2018BacksteppingNonlinear}, more advanced methods that achieve exponential stability have been proposed such as the Adaptive Robust Control (ARC) method \cite{Yao1997}.
Based on the adaptive backstepping technique and upgraded with a robust compensation term to address system disturbances and unmodeled nonlinearities, the ARC method was initially established for a class of nonlinear SISO systems \cite{Yao1997}, proving usefulness in various applications. Generalization of this method for multiple-input multiple-output (MIMO) systems in the semi-strict feedback form was formulated later on \cite{Yao2001MIMO}.
\par
\par
One of the main shortcomings of the backstepping design process is the explosion of complexity when dealing with systems of high-order dynamics \cite{Swaroop2000}, where the calculation of certain terms becomes prohibitive in systems with order higher than three \cite{Dong2012CommandFilteredBackstepping}.
Thus, smoothing functions are exploited \cite{Wen2011RAC_nonlinear} and filters are introduced at the virtual commands \cite{Dong2012CommandFilteredBackstepping}, which is further improved by using the finite-time control technique \cite{Yu2018BacksteppingNonlinear}.
Unfortunately, even though filters can solve the explosion of terms' problem, they introduce time delays in the signals, which may require the filtration of each component of the virtual commands \cite{Dong2012CommandFilteredBackstepping}.
\par
Researcher have formulated the backstepping control law for different classes of systems in a PID-like structures. For instance, this analogy was formulated for linear second-order minimal phase systems \cite{Benaskeur2002_BacksteppingBasedPID}.
In \cite{Skjetne2004_IntegralControlBackstepping}, integral control backstepping was studied for a class of nonlinear mechanical systems of relative degree two, and one formulation of the control law in a PID-like structure was presented, while showing the drawbacks of the design.
A similar design was applied to quadrotor Unmanned Aerial Vehicles (UAVs), but with increased number of tuning parameters in \cite{Mian2008_BacksteppingBasedPID}, and for a linearized version of the system in \cite{Kartal2019Backstepping_PID_UAV}.
\par
This work is an attempt to uncover an underlying link between Lyapunov stability theory through the adaptive backstepping control technique, and the nonlinear PID control structure for second-order systems, which cover a wide spectrum of real-life practical systems \cite{Pan2010Robust2ndOrderLyapunov}.
In this work, we formulate the adaptive backstepping control method for a class of second-order nonlinear systems in a compact two-degrees-of-freedom (DOF) form, which has a PID feedback component and a feedforward model compensation component, then draw the similarity link between the backstepping controller gains and the PID gains with a minimal number of dependencies. This proposition opens the door to use well-established PID tuning rules to tackle the problem of tuning Lyapunov-based adaptive backstepping controllers. Another benefit of this formulation is to digest the contents of the derivatives of the virtual commands to show their true form, allowing a more prudent employment of filters, thus reducing excessive time-delays.
\par
This effort differs from existing work on this topic by formulating the PID similarity with a minimal number of parameters and tuning gains \cite{Benaskeur2002_BacksteppingBasedPID,Mian2008_BacksteppingBasedPID}, and a minimal number of dependencies between the tuning gains \cite{Skjetne2004_IntegralControlBackstepping,Kartal2019Backstepping_PID_UAV}, while explicitly analysing the influence of gains selection on the similarity and system stability.
The proposed formulation is validated on a quadrotor UAV with variable payload, subjected to external disturbances.
\par
The rest of this paper provides the problem formulation in Section~\ref{sec_problem_formulation}, followed by the controller design in Section \ref{sec_control_system_design}. The resulting control law is reformulated and discussed in Section~\ref{sec_PID_form}, and Section \ref{sec_sim_results} presents a brief description of the quadrotor dynamics and the simulation and experimental results to validate the performance of the proposed control system design. Section \ref{sec_conclusion} concludes the paper and provides an outlook into future work.
%
\section{Problem Formulation}
\label{sec_problem_formulation}
\subsection{Preliminaries} \label{sec_preliminaries}
We first introduce some notations that are used throughout this paper. We let the set of positive-real numbers $\{x \in \mathbb{R}\,|\,x>0 \}$ be denoted as $\mathbb{R}_{>0}$, and the set of non-negative real numbers $\{x \in \mathbb{R}\,|\,x\geq 0 \}$ be denoted as $\mathbb{R}_{\geq0}$.
Also, let $s_{\bigcdot}$, $c_{\bigcdot}$, and $t_{\bigcdot}$ respectively be the sine, cosine, and tangent functions for some angle ($\bigcdot$).
In addition, let $\norm{\cdot}$ denote the $L_2$ norm of a signal, and for some estimate $\hat{x}$ of $x \in [x_{\mathrm{min}},x_{\mathrm{max}}]$, $\mathrm{Proj}_{x}(\cdot)$ is a projection function defined as \cite{Hu2010DIARC_Gantry}:
\begin{equation} \label{eq_projection_function}
\mathrm{Proj}_{\hat{x}}(\cdot)=
    \begin{cases}
        0 & \text{if $\hat{x} = x_{\mathrm{max}}$ and $\cdot > 0$}\\
        0 & \text{if $\hat{x} = x_{\mathrm{min}}$ \,and $\cdot < 0$}\\
        \,\cdot & \text{otherwise}.\\
    \end{cases}
\end{equation}
\subsection{Problem Formulation}
Consider a second-order nonlinear time-varying system with state vectors $\bm{X}_1=[x_1, ..., x_n]^{\intercal}$ and $\bm{X}_2=[\dot{x}_1, ..., \dot{x}_n]^{\intercal}$, and control input vector $\bm{U}=[u_1, ..., u_n]^{\intercal} \in \mathbb{R}^{n}$, $n \in \mathbb{N}^+$, of the form:
\begin{equation} \label{eq_ss_model}
\begin{split}
    \dot{\bm{X}}_1 & = \bm{X}_2, \\
    \dot{\bm{X}}_2 & = \bm{f}(\bar{\bm{X}}) + \bm{\phi} (\bar{\bm{X}}) \bm{\theta}(t) + \bm{g}(\bar{\bm{X}},t)\bm{U} + \bm{\Delta}(t),
\end{split}
\end{equation}
\noindent where $\bar{\bm{X}}=\{\bm{X}_1,\bm{X}_2\}$, $\bm{f} \in \mathbb{R}^n$ represents known nonlinear functions vector, $\bm{\phi} \in \mathbb{R}^{n\times l}$ represents known nonlinear regressors matrix with $l \in \mathbb{N}^+$ being the number of unknown linear system parameters, grouped in the vector $\bm{\theta} \in \mathbb{R}^{l}$, $\bm{g}  \in \mathbb{R}^{n \times n}$ represents input-multiplied unknown nonlinear functions, and the vector $\bm{\Delta}=[\Delta_1,...,\Delta_n]^{\intercal} \in  \mathbb{R}^{n}$ represents the unmodeled nonlinearities and the time-varying external disturbances.
The controller's objective is to follow a reference command $\bm{X}_{1\mathrm{d}}(t)=[x_{1\mathrm{d}},...,x_{n\mathrm{d}}]^{\intercal}$.
\begin{assumption}\label{assump_smooth_bounded_Xd}
    $\bm{X}_{1\mathrm{d}}(t)$, along with its first and second time derivatives, are assumed to be smooth and bounded to maintain the boundedness of all states and signals in system.
\end{assumption}
\begin{assumption}\label{assump_bounded_functions}
  If $\bm{X}_2$ is bounded, then the functions $\bm{f}$, $\bm{\phi}$, and $\bm{g}$, and their first-order and second-order derivatives with respect to $\bm{X}_2$ are bounded.
\end{assumption}
\begin{assumption}\label{assump_bounded_uncert_dist}
  The parametric uncertainties and the unmodeled nonlinearities and external disturbances term belong to known sets such that:
  \begin{equation}
      \begin{split}
          \theta_{i,\mathrm{min}} & \leq \theta_i \leq \theta_{i,\mathrm{max}}, \; i=1,...,l, \\
          g_{ij,\mathrm{min}} & \leq g_{ij} \leq g_{ij,\mathrm{max}}, \; i,j=1,...,n,\quad
          \norm{\Delta} \leq \bar{\Delta},
      \end{split}
  \end{equation}
 to ensure boundedness, where $\theta_i$ and $g_{ij}$ are elements of $\bm{\theta}$ and $\bm{g}$ respectively, and $\bar{\Delta}$ is a positive constant.
\end{assumption}
%
\section{Control System Design} \label{sec_control_system_design}
\subsection{Parameter Estimation Law}
In practice, since the system parameters are time-varying and unknown external disturbances exist, $\bm{\theta}(t)$ and $\bm{g}(\bar{\bm{X}},t)$ are substituted with their estimates, $\hat{\bm{\theta}}(t)$ and $\hat{\bm{g}}(t)$, respectively, in the controller design. 
A recursive least squares (RLS) adaptation law with projection mapping is adopted for the proposed controller architecture \cite{Yao2002DIARC}. 
Note that $\hat{\bm{g}}(t)$ must remain invertible.
\par
%
\subsection{Controller Design}
The in brackets functions dependencies will be dropped for ease of reading purpose.
The error dynamics vector is defined as:
    $\bm{e}_1=\bm{X}_1-\bm{X}_{1\mathrm{d}}$.
For a second-order system, the backstepping process involves two steps. By referring to the error dynamics, we define a candidate Lyapunov function, $\mathcal{V}_1=\frac{1}{2}\bm{e}_1^{\intercal}\bm{e}_1$, yielding the following derivative:
    $\dot{\mathcal{V}}_1=\bm{e}_1^{\intercal}\dot{\bm{e}}_1=\bm{e}_1^{\intercal}(\dot{\bm{X}_1}-\dot{\bm{X}}_{1\mathrm{d}})$. 
%
To stabilize $\bm{e_1}$, the virtual control input $\bm{\alpha} \in \mathbb{R}^n$ replaces $\dot{\bm{X}}_1$ such that:
    $\bm{\alpha}=\dot{\bm{X}}_{1\mathrm{d}} - \bm{k}_1 \bm{e}_1$,
where $\bm{k}_1 \in \mathbb{R}^{n\times n}_{>0}$ is a diagonal matrix.\
The second step entails defining another error term that relates the virtual input to the state derivatives as:
$\bm{e_2}=\bm{X}_{2} - \bm{\alpha}$.
%
Differentiating $\bm{e_2}$, and referring to (\ref{eq_ss_model}) while including the time-varying parameters and unknown external disturbances' effect, yields the following error dynamics:
\begin{equation} \label{eq_e2_dot_modified}
    \dot{\bm{e}}_2 = \bm{f} + \bm{\phi} \hat{\bm{\theta}} + \hat{\bm{g}}\bm{U} - \dot{\bm{\alpha}} + \bm{d},
\end{equation}
\noindent where $\bm{d}=[d_1,...,d_n]^{\intercal}$ is regarded as the lumped mismatch in the error compensation due to estimation error, unmodeled nonlinearities, and disturbances.
Note that $\dot{\bm{e}}_2$ explicitly includes the control input vector $\bm{U}$.
The term $\bm{d}$ can be expressed as the combination of a low-frequency component, $\bm{d}_{\mathrm{c}}$, and a high-frequency component, $\tilde{\bm{d}}^{\ast}$, such that:
\begin{equation} \label{eq_d_nl_dist}
    \bm{d} := \bm{d}_{\mathrm{c}} + \tilde{\bm{d}}^{\ast} = -\bm{\phi} \tilde{\bm{\theta}} - \tilde{\bm{g}} \bm{U} + \bm{\Delta},
\end{equation}
\noindent where $\tilde{\bm{\theta}}=\hat{\bm{\theta}} - \bm{\theta}$ and $\tilde{\bm{g}}=\hat{\bm{g}} - \bm{g}$.
\begin{remark}\label{remark_bounded_uncertainties}
    By Assumption \ref{assump_bounded_uncert_dist}, system uncertainties and disturbances are bounded, such that $\norm{\bm{d}} \leq \bar{d}$, where $\bar{d} \in \mathbb{R}_{\geq0}$.
\end{remark}
The proposed control law is of the following form:
\begin{equation} \label{eq_U_adaptive_robust}
    \begin{split}
        \bm{U} & = \hat{\bm{g}}^{-1}(-\bm{e}_1 + \dot{\bm{\alpha}} - \bm{f} - \bm{\phi} \hat{\bm{\theta}} - \bm{k}_2 \bm{e}_2 - \hat{\bm{d}}_{\mathrm{c}} + \bm{u}_{\mathrm{r}}),\\   
        \dot{\hat{\bm{d}}}_{\mathrm{c}} & = \mathrm{Proj}_{\hat{\bm{d}}_{\mathrm{c}}} (\bm{\gamma} \bm{e}_{2}), \qquad
        \bm{u}_{\mathrm{r}}  = -\frac{1}{4 \epsilon} h^2 \bm{e}_2,
    \end{split}
\end{equation}
where $\mathrm{Proj}_{\hat{\bm{d}}_{\mathrm{c}}}(\cdot)=[\mathrm{Proj}_{\hat{d}_{\mathrm{c}1}}(\cdot_1),...,\mathrm{Proj}_{\hat{d}_{\mathrm{c}n}}(\cdot_n)]^{\intercal}$, $\bm{\gamma} \in \mathbb{R}^{n \times n}_{\geq 0}$ is a diagonal matrix affecting the adaptation speed, $\epsilon \in \mathbb{R}_{>0}$ is a design parameter that quantifies the level of error attenuation, and $h$ is a smooth function chosen such that $h \geq \norm{\bm{\phi}} \norm{\bm{\theta}_{\mathrm{max}} - \bm{\theta}_{\mathrm{min}}} + \norm{\bm{U}} \norm{\bm{g}_{\mathrm{max}}-\bm{g}_{\mathrm{min}}}  + \norm{\bar{\bm{\Delta}}}$.
Note that $\mathrm{Proj}_{\hat{\bm{d}}}(\cdot)$ limits $\hat{\bm{d}}_{\mathrm{c}}$ to the set $\Omega_{\bm{d}_{\mathrm{c}}} \triangleq \{\hat{\bm{d}}_{\mathrm{c}} : -\bar{d} \leq \hat{\bm{d}}_{\mathrm{c}i} \leq \bar{d}, i=\{1,...,n\}\}$.
\begin{theorem}\label{theorem_adaptive_controller}
    Consider the system described in (\ref{eq_ss_model}) with parametric uncertainties, unmodeled nonlinearities, and unknown disturbances, and suppose that Assumptions~\ref{assump_smooth_bounded_Xd}, \ref{assump_bounded_functions}, and \ref{assump_bounded_uncert_dist} hold.
    The control law (\ref{eq_U_adaptive_robust}) guarantees the stability of this system for a set of gains $\bm{k}_1$ and $\bm{k}_2$, and ensures that all signals in the system are bounded and the tracking error has a transient performance and the steady-state accuracy governed by the choice of $\epsilon$.
\end{theorem}
\begin{IEEEproof}
    Choose the following augmented Lyapunov function, $\mathcal{V}_2$, apply differentiation, and substitute $\bm{U}$ from (\ref{eq_U_adaptive_robust}) and $\bm{d}$ from (\ref{eq_d_nl_dist}) in (\ref{eq_e2_dot_modified}), we get:
        \begin{align*}
            &\mathcal{V}_2 = \frac{1}{2} \bm{e}_1^{\intercal}\bm{e}_1+\frac{1}{2} \bm{e}_2^{\intercal}\bm{e}_2 +\frac{1}{2}  \bm{\tilde{d}}_{\mathrm{c}}^{\intercal} \bm{\gamma}^{-1} \bm{\tilde{d}}_{\mathrm{c}},\\
            & \dot{\mathcal{V}}_2  =\scalebox{0.90}{$  - \bm{e}_1^{\intercal} \bm{k}_1 \bm{e}_1 - \bm{e}_2^{\intercal} \bm{k}_2 \bm{e}_2 + \bm{e}_2^{\intercal} (\bm{d}_{\mathrm{c}}-\hat{\bm{d}}_{\mathrm{c}} + \tilde{\bm{d}}^{\ast} +\bm{u}_{\mathrm{r}})  +  \bm{\tilde{d}}_{\mathrm{c}}^{\intercal} \bm{\gamma}^{-1} \dot{\bm{\hat{d}}}_{\mathrm{c}} $},
        \end{align*}
    %
    where $\bm{\tilde{d}}_{\mathrm{c}} = \bm{\hat{d}}_{\mathrm{c}} - \bm{d}_{\mathrm{c}}$; note that since $\bm{d}_{\mathrm{c}}$ is considered the low-frequency component of $\bm{d}$, we can safely assume that $\dot{\bm{d}}_{\mathrm{c}} = 0$.
    By substituting $\dot{\hat{\bm{d}}}_{\mathrm{c}}$ from (\ref{eq_U_adaptive_robust}) in $\dot{\mathcal{V}}_2$, we get: $\dot{\mathcal{V}}_{2}=- \bm{e}_1^{\intercal} \bm{k}_1 \bm{e}_1 - \bm{e}_2^{\intercal} \bm{k}_2 \bm{e}_2 + \bm{e}_2^{\intercal} (\tilde{\bm{d}}^{\ast}+\bm{u}_{\mathrm{r}})$.
    Still, the robust feedback term, $\bm{u}_{\mathrm{r}}$, should
    satisfy the following two robust performance conditions \cite{Yao2001MIMO}:
    \begin{equation} \label{eq_ur_cond}
            \bm{e}_2^{\intercal} (\bm{u}_{\mathrm{r}}- \tilde{\bm{d}}_{\mathrm{c}} + \tilde{\bm{d}}^{\ast})  \leq \bm{\epsilon}, \qquad
            \bm{e}_2^{\intercal} \bm{u}_{\mathrm{r}} \leq 0.\quad
    \end{equation}
    By choosing $\bm{u}_{\mathrm{r}}$ as described in (\ref{eq_U_adaptive_robust}), the conditions in (\ref{eq_ur_cond}) are satisfied and robust performance is guaranteed \cite{Yao2001MIMO}.
\end{IEEEproof}

\section{Expressing the Control Law in PID Form} \label{sec_PID_form}
In practice, the nonlinear robust gain, $\frac{1}{4 \epsilon} h^2$,can be substituted by a large-enough positive constant, $\bm{k}_{2,0}$, to simplify the control law. 
This will make the effect of $\bm{u}_{\mathrm{r}}$ similar to that of the linear term $-\bm{k}_2 \bm{e}_2$. Thus by choosing $\bm{k}_2 \succ_{\mathbb{R}^{n \times n}_{\geq 0}} \bm{k}_{2,0}$, we can still guarantee a practical robust performance within a large-enough working range \cite{Hu2010DIARC_Gantry}.
Given this modification, the control law (\ref{eq_U_adaptive_robust}) can be expanded and simplified by substituting $\bm{e}_1$, $\bm{\alpha}$, and $\bm{e}_2$ into (\ref{eq_U_adaptive_robust}):
\begin{equation} \label{eq_U_ABS_to_PID}
    \begin{split}
        \bm{U} & \approx \bm{U}' = \hat{\bm{g}}^{-1}\big[-\bm{e}_1 + (\ddot{\bm{X}}_{1\mathrm{d}} - \bm{k}_1\dot{\bm{e}}_1) - \bm{f} - \bm{\phi}  \hat{\bm{\theta}} - \\
        & \qquad \quad \;\, - \bm{k}_2(\dot{\bm{X}}_1 - (\dot{\bm{X}}_{1\mathrm{d}} - \bm{k}_1 \bm{e}_1)) - \hat{\bm{d}}_{\mathrm{c}} \, \big],\\ 
        \dot{\hat{\bm{d}}}_{\mathrm{c}} & = \mathrm{Proj}_{\hat{\bm{d}}_{\mathrm{c}}} \big(\bm{\gamma} \big[\dot{\bm{X}}_1-(\dot{\bm{X}}_{\mathrm{1d}} - \bm{k}_1 \bm{e}_1)\big]\big),\\
    \end{split}
\end{equation}
\noindent where $\dot{\bm{e}}_1=\dot{\bm{X}}_1-\dot{\bm{X}}_{1\mathrm{d}}$. Thus, we can express (\ref{eq_U_ABS_to_PID}) as a two-DOF controller with a PID feedback component and a feedforward model compensation term, dubbed here as Integrated Control System (ICS):
\begin{equation} \label{eq_U_PID}
    \begin{split}
        & \bm{U}'= \scalebox{0.95}{$\hat{\bm{g}}^{-1}\big[  -\bm{k}_{\mathrm{P}}\bm{e}_1 -\bm{k}_{\mathrm{D}} \dot{\bm{e}}_1-\bm{k}_{\mathrm{I}}\bm{e}_1^{\mathrm{I}} + \ddot{\bm{X}}_{1\mathrm{d}} -\bm{f} -\bm{\phi}  \hat{\bm{\theta}} \big]$},\\
        & \dot{\bm{e}}_1^{\mathrm{I}}=\mathrm{Proj}_{\bm{e}_1^{\mathrm{I}}} (\bm{e}_1+\bm{k}_1^{-1}\dot{\bm{e}}_1),
        \end{split}
\end{equation}
\noindent where $\bm{I}_n \in  \mathbb{R}^{n \times n}$ is the identity matrix and
\begin{equation} \label{eq_K_PID}
        \bm{k}_{\mathrm{P}} =\bm{I}_n+\bm{k}_1 \bm{k}_2, \quad
        \bm{k}_{\mathrm{D}} =\bm{k}_1+\bm{k}_2, \quad
        \bm{k}_{\mathrm{I}} =\bm{\gamma k}_1.
\end{equation}
The first three terms in (\ref{eq_U_PID}) represent an error-based PID feedback component with gains $\bm{k}_{\mathrm{P}}$, $\bm{k}_{\mathrm{D}}$ and $\bm{k}_{\mathrm{I}}$, and the last three terms represent a feedforward component based on desired accelerations and model compensation.
\par
To provide a clear interpretation of the control law, we assume that the integral term does not saturate, thus after integrating the term $\bm{k}_1^{-1}\dot{e}_1$, the control law in (\ref{eq_U_PID}) and (\ref{eq_K_PID}) incurs the following changes:
\begin{equation} \label{eq_adjust_PID}
        \bm{k}_{\mathrm{P}} =\bm{I}_n+\bm{k}_1 \bm{k}_2 + \bm{\gamma}, \qquad
        \dot{\bm{e}}_1^{\mathrm{I}}=\bm{e}_1.
\end{equation}
\par
\begin{figure} 
\centerline{\includegraphics[width=3.34in]{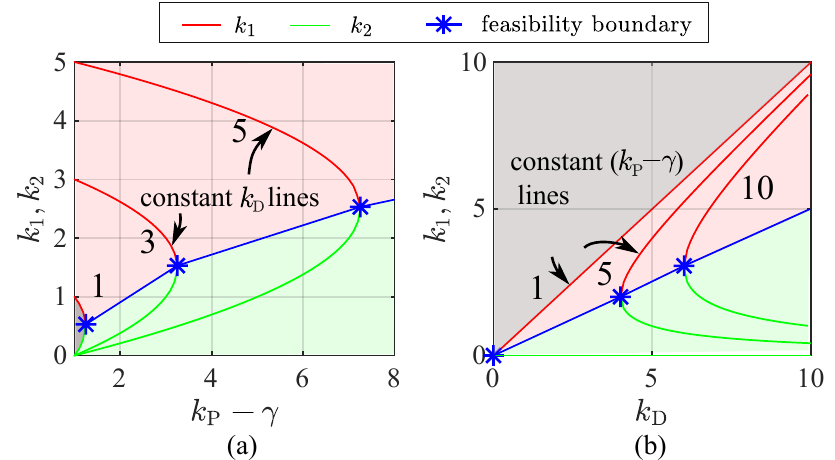}}
\caption{Coupled $k_1$ and $k_2$ for specific $k_{\mathrm{P}}$ and $k_{\mathrm{D}}$ values: (a) at constant $k_{\mathrm{D}}$ and (b) at constant $(k_{\mathrm{P}}-\gamma)$. The red and the green regions represent the sets of all feasible $k_1$ and $k_2$ combinations respectively.}
\label{fig_K12_vs_Kpd}
\end{figure}

Let $k_{\mathrm{P}}$, $k_{\mathrm{D}}$, $k_1$, $k_2$, and $\gamma$, represent the diagonal elements of an arbitrary single row of $\bm{k}_{\mathrm{P}}$, $\bm{k}_{\mathrm{D}}$, $\bm{k}_1$, and $\bm{k}_2$, respectively. 
A set of $k_1$ and $k_2$ gains can be determined by solving the following second-order equation, derived from the definitions of $\bm{k}_{\mathrm{P}}$ and $\bm{k}_{\mathrm{D}}$, for any desired values of $k_{\mathrm{P}}$ and $k_{\mathrm{D}}$:
\begin{equation} \label{eq_K12_from_Kpd}
    \begin{split}
        & k_{1,2}^2 + (-k_{\mathrm{D}})k_{1,2} + (k_{\mathrm{P}}-\gamma-1) = 0,\\
        & k_1:=\mathrm{max}\{k_{1,2}\}, \quad
        k_2:=\mathrm{min}\{k_{1,2}\},
    \end{split}
\end{equation}
\noindent where $k_{1,2}$ are the roots of the equation.
An implementation of (\ref{eq_K12_from_Kpd}) is shown in Fig.~\ref{fig_K12_vs_Kpd} for various $k_{\mathrm{P}}$ and $k_{\mathrm{D}}$ values under the constraint of a constant $k_{\mathrm{D}}$ in Fig.~\ref{fig_K12_vs_Kpd}(a) and a constant $k_{\mathrm{P}}$ in Fig.~\ref{fig_K12_vs_Kpd}(b). The blue asterisks correspond to the last feasible solution of (\ref{eq_K12_from_Kpd}) for each respective streamline (constant $k_{\mathrm{P}}$ or $k_{\mathrm{D}}$), which is the maximum in Fig.~\ref{fig_K12_vs_Kpd}(a) and the minimum in Fig.~\ref{fig_K12_vs_Kpd}(b).
\par
This is an important finding since it facilitates the design and tuning process of the robust adaptive controller gains, and allows the application of well-established PID tuning methods. 
Additionally, unfeasible PD gains that result in non-real or non-positive $k_{1,2}$ roots can be directly excluded. Furthermore, the designer can gain insight into the performance perspective of the adaptive backstepping control law as the gains vary, in addition to the traditional stability perspective.
%
%
\begin{remark} \label{remark_U_PID_advantages}
    Tuning the gains $k_1$ and $k_2$ does not directly influence the system performance as can be universally understood and expected from PD gains. Eventually, this can mislead the control engineer because of the dual effect that each one of the gains $k_1$ and $k_2$ has on both $k_{\mathrm{P}}$ and $k_{\mathrm{D}}$, as shown in Fig.~\ref{fig_K12_vs_Kpd}. It can also be seen that not every combination of $k_{\mathrm{P}}$ and $k_{\mathrm{D}}$ gains can be obtained from the backstepping control law (\ref{eq_U_PID}), where for a specific $k_{\mathrm{D}}$ value, there is a maximum possible $k_{\mathrm{P}}$ value (see Fig.~\ref{fig_K12_vs_Kpd}(a)) for which the system is stable:
    \begin{equation} \label{eq_kp_max}
      k_{\mathrm{P},\mathrm{max}}=\frac{k_{\mathrm{D}}^2}{4}+1+\gamma.
    \end{equation}
    Alternatively, it can be stated that for a specific $k_{\mathrm{P}}$ value, there is a minimum possible $k_{\mathrm{D}}$ value (see Fig.~\ref{fig_K12_vs_Kpd}(b)):
    \begin{equation} \label{eq_kD_min}
        k_{\mathrm{D},\mathrm{min}}=2\sqrt{k_{\mathrm{P}}-\gamma-1}.
    \end{equation}    
\end{remark}

\begin{remark} \label{remark_gains_tuning}
    To determine the gains $\bm{k}_1$ and $\bm{k}_2$ of the control law (\ref{eq_U_adaptive_robust}), $\bm{k}_{\mathrm{P}}$, $\bm{k}_{\mathrm{D}}$, and $\bm{k}_{\mathrm{I}}$ are obtained via PID tuning based on desired performance. The resulting PD gains must give a feasible solution, that is real and positive, when plugged in (\ref{eq_K12_from_Kpd}); otherwise, the requirements of Theorem \ref{theorem_adaptive_controller} are not met.   
\end{remark}

\begin{remark} \label{remark_selective_filtration_in_U_PID}
    Since $\dot{\bm{e}}_1$ is a differentiated quantity, filtration can be applied to attenuate the signal noise. However, $\dot{\bm{e}}_1$ is hidden inside $\bm{e}_2$ and $\dot{\bm{\alpha}}$ in the adaptive robust control law (\ref{eq_U_adaptive_robust}), thus it is remarkably easier to design an appropriate filter when the control law is written in the PID-like form (\ref{eq_U_PID}) to prevent excessive filtration leading to loss of information and time delays as in \cite{Dong2012CommandFilteredBackstepping}. This does not only address a potential shortcoming in the ARC design, but in the generic backstepping controller design as well.
\end{remark}
\begin{remark} \label{remark_Kpd_effect_on_K12}
    For fine-tuning purposes, if the effect of varying the embedded $k_{\mathrm{P}}$ ($k_{\mathrm{D}}$) gain of the control law (\ref{eq_U_adaptive_robust}) is desired, then $k_1$ and $k_2$ gains should be tuned according to Fig.~\ref{fig_K12_vs_Kpd}(a) (Fig.~\ref{fig_K12_vs_Kpd}(b)).
\end{remark}
%
\section{Simulation Example}
\label{sec_sim_results}
\begin{figure}
\centerline{\includegraphics[width=3.34in]{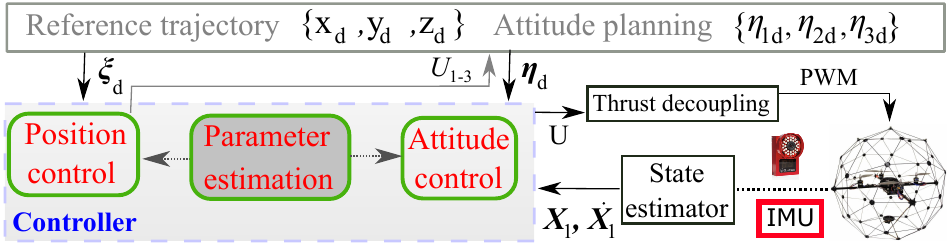}}
\caption{Control system architecture.}
\label{fig_controller_diagram}
\end{figure}

Among many possible applications, the ICS is validated on a quadrotor UAV with variable payload,
using the controller structure shown in Fig. \ref{fig_controller_diagram}, which has an inner-loop (attitude) and outer-loop (position) controller configuration.
%
\subsection{Dynamic Model} \label{sec_quadrotor_model}
\begin{figure}
\centerline{\includegraphics[width=3.34in]{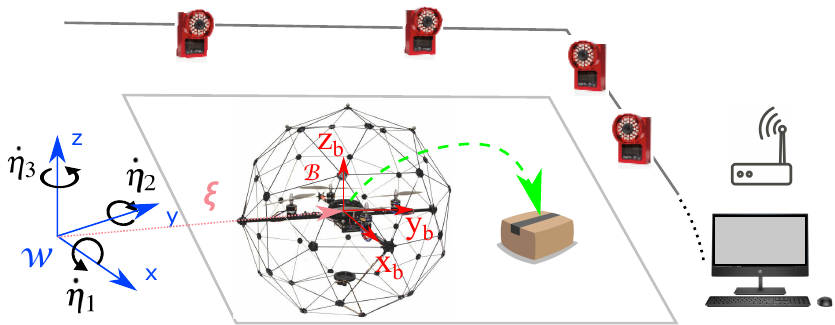}}
\caption{Quadrotor system configuration with an indoor motion capture system.}
\label{fig_quadrotor_annotations}
\end{figure}

%
Consider the quadrotor depicted in Fig.~\ref{fig_quadrotor_annotations}. Let $\mathcal{W}$ be the inertial frame of reference, and $\mathcal{B}$ be the body-fixed frame at the quadrotor's centroid, $\mathcal{O}_{B}$. The quadrotor has a mass $m \in \mathbb{R}_{>0}$ and an inertia tensor $\bm{J} \in \mathbb{R}^{3 \times 3}$ with respect to $\mathcal{O}_{B}$. 
Let $\bm{r}=[r_x,r_y,r_z]^{\intercal} \in \mathbb{R}^3$ represents the quadrotor's center of mass (CoM) coordinates in $\mathcal{B}$.
The vector $\bm{\xi}=[\mathrm{x},\mathrm{y},\mathrm{z}]^\intercal$ represents the quadrotor's translational position in $\mathcal{W}$, and $\bm{\eta}=[\eta_1,\eta_2,\eta_3]^{\intercal}$ represents the vector of the Euler angles for roll, pitch, and yaw motions, respectively.
\par
The quadrotor's dynamical model is described as follows:
\begin{equation} \label{eq_dyn_model}
 \begin{split} 
    m\ddot{\bm{\xi}} & =  -m\bm{G} + \mathbf{R}_\mathrm{t} \bm{F}_{\mathcal{B}} +\Delta_{\xi}, \\
    \bm{J} \mathbf{R}_\mathrm{r} \bm{\ddot{\eta}} & = \bm{\tau}_{\mathcal{B}} + \bm{r} \times \bm{F}_{\mathcal{B}} + \Delta_{\eta},   
 \end{split}
\end{equation}
\noindent where $\bm{G}=[0,0,\textsl{g}]^{\intercal}$ is the gravitational acceleration vector with $\textsl{g} = \SI[unitsep=medium]{9.81}{\meter\per\second\squared}$ being the gravity constant; $\mathbf{R}_\mathrm{t}$ and $\mathbf{R}_\mathrm{r}$ are the translational and rotational velocity transformation matrices between $\mathcal{W}$ and $\mathcal{B}$ \cite{Kourani2018}; $\bm{F}_{\mathcal{B}}=[0 , 0 , F_{\mathrm{t}}]^{\intercal} \in \mathbb{R}^3$ is the rotors' total force in $\mathcal{B}$, with $F_{\mathrm{t}} = K_{\mathrm{t}} \sum_{i=1}^4{\omega_i^2}$; $\bm{\tau}_{\mathcal{B}}=[T_1,T_2,T_3]^{\intercal}=[K_{\mathrm{t}} L(\omega_3^2-\omega_4^2) ,\; K_{\mathrm{t}} L(\omega_1^2-\omega_2^2) ,\; K_{\mathrm{Q}} (\omega_1^2+\omega_2^2-\omega_3^2-\omega_4^2)]^{\intercal}$ is the vector of rotor torques, with $K_{\mathrm{t}}$ and $K_{\mathrm{Q}}$ being the rotor thrust and torque constants, respectively, $L$ is the arm length, and $\omega_i$ is the $i^{th}$ rotor angular speed with $i\in[1,2,3,4]$. The control inputs, $\omega_i$, are computed from the above equations as done in \cite{Kourani2018}. 
The terms $\Delta_{\xi}$ and $\Delta_{\eta}$ $\in \mathbb{R}^3$ are the unmodeled nonlinearities and external disturbances \cite{Kourani2018,Zhang2019}.
\par
\subsection{Problem Formulation}
The quadrotor dynamics can be written in the form of (\ref{eq_ss_model}) by choosing:
\begin{equation} \label{eq_quadrotor_ss_model}
    \begin{split}
        \bm{X}_1 & =[\mathrm{x},\mathrm{y},\mathrm{z}, \eta_1, \eta_2, \eta_3]^{\intercal}, \;
        \bm{X}_2 =[\mathrm{\dot{x}},\mathrm{\dot{y}},\mathrm{\dot{z}}, \dot{\eta}_1, \dot{\eta}_2, \dot{\eta}_3]^{\intercal},
         \\
        \bm{f} & =[0,0,-\textsl{g},0,0,0]^{\intercal}, \;\;\quad
        \bm{\phi} = [\bm{O}_{3 \times 2}; \bm{\phi}_2],\\
        \bm{U} & = [\bm{F}_{\mathcal{W}};\bm{\tau}_{\mathcal{B}}], \qquad \qquad \quad    
        \bm{\theta} = [r_x, r_y]^{\intercal}, \\
        \bm{g} & =[m^{-1} \bm{I}_3, \bm{O}_3; \bm{O}_3 , (\bm{J} \mathbf{R}_\mathrm{r} )^{-1}], \quad \bm{\Delta} =[\Delta_{\xi};\Delta_{\eta}],
    \end{split}
\end{equation}
\noindent where $\bm{O}_3 \in  \mathbb{R}^{3 \times 3}$ is the null matrix, and the elements $\bm{\theta}$, $\bm{g}$, and $\bm{\Delta}$ represent the unknown entities in the system.  
Acrobatic maneuvers are excluded, such that the singular point of $\eta_3=\frac{\pm\pi}{2}$ is not reached and $\bm{g}$ remains Lipschitz continuous. The matrix $\bm{\phi}_2 \in \mathbb{R}^{3\times2}$ is defined as:
$\bm{\phi}_2 = [0,\, -F_{\mathrm{t}} c_{\eta_2}/J_{xx,0};\, F_{\mathrm{t}}/J_{yy,0},\, -F_{\mathrm{t}} t_{\eta_1} s_{\eta_2}/J_{xx,0};\, 0,\, F_{\mathrm{t}} s_{\eta_2} / \, \allowbreak (c_{\eta_1} J_{xx,0})]$, where $J_{xx,0}$ and $J_{yy,0}$ are respectively the nominal values of $J_{xx}$ and $J_{yy}$.
The translational control input vector is expressed as:
\begin{align*}
\bm{F}_{\mathcal{W}}=
[s_{\eta_1}c_{\eta_2}s_{\eta_3}+s_{\eta_2}c_{\eta_3};\,
        s_{\eta_2}s_{\eta_3}-s_{\eta_1}c_{\eta_2}c_{\eta_3};\,
        c_{\eta_1}c_{\eta_2}] F_{\mathrm{t}}.
\end{align*}        
%
\subsection{Reference Signals and Control Inputs}
The controller's goal is to have the state vector, $\bm{X}_1$, track a desired state vector, $\bm{X}_{1\mathrm{d}}=[\bm{\xi}_{\mathrm{d}};\bm{\eta}_{\mathrm{d}}]=[\mathrm{x}_{\mathrm{d}},\mathrm{y}_{\mathrm{d}},\mathrm{z}_{\mathrm{d}}, \eta_{1\mathrm{d}},\eta_{2\mathrm{d}},\eta_{3\mathrm{d}}]^{\intercal}$ by using the control input $\mathrm{U}=[F_{\mathrm{tc}},T_{1\mathrm{c}},T_{2\mathrm{c}},T_{3\mathrm{c}}]^{\intercal}$, where the subscript $(\bigcdot)_{\mathrm{c}}$ refers to the command input.
To reduce complexity, the quadrotor's motion control system is divided into two subsystems: an inner-loop that controls rotations and an outer-loop that controls position. Using  $[u_1,u_2,u_3]$ that are generated by the position controller, we calculate the total thrust command as:
\begin{equation} \label{eq:u_1}
    F_{\mathrm{tc}}=\sqrt{u_1^2+u_2^2+u_3^2}.
\end{equation}
The desired attitude angles, $\eta_{1\mathrm{d}}$ and $\eta_{2\mathrm{d}}$, are computed to properly orient the quadrotor and satisfy $[u_1,u_2,u_3]$. Thus, from $\bm{F}_{\mathcal{W}}$ we get:
\begin{equation} \label{eq_etas}
 \begin{split}
    & \eta_{1{\mathrm{d}}}=\text{arctan}\big((u_1 s_{\eta_3} -u_2 c_{\eta_3})/u_3\big),\\
    & \eta_{2{\mathrm{d}}}=\text{arcsin}\big((u_1 c_{\eta_3}+u_2 s_{\eta_3})/F_{\mathrm{tc}}\big).
 \end{split}
\end{equation}
This gives a fully-defined desired state vector, $\bm{X}_{1\mathrm{d}}$, which is to be tracked by $\bm{X}_1$.

\subsection{Simulation Model Elements}
\subsubsection{Simulation settings}
The proposed control and estimation scheme is simulated in the MATLAB/Simulink\,\textsuperscript{\tiny\textregistered} environment, with a mission performed on the derived high-fidelity nonlinear dynamic model, which is augmented to include the propeller motors model, force and torque disturbances (wind gust of a magnitude of $\SI[unitsep=medium]{5}{m/s}$), feedback delays, and induced sensor noise. Actual measurement signals were conditioned and fused via a Kalman filter to produce smooth estimates of the recorded measurements, and the estimation algorithm is designed using the RLS method.
\subsubsection{Quadrotor specifications}
The quadrotor model is based on the Quanser QBall-2 platform, which has the following specifications: $m=\SI[unitsep=medium]{1.76}{\kilogram}$, $J_{xx}=J_{yy}=\SI[unitsep=medium]{0.03}{\kilogram \square\meter}$, $J_{zz}=\SI[unitsep=medium]{0.04}{\kilogram \square\meter}$, $L=\SI[unitsep=medium]{0.2}{\meter}$, $K_{\mathrm{t}} = \SI[unitsep=medium]{13}{\newton}$, and $K_{\mathrm{Q}}=\SI[unitsep=medium]{0.4}{\newton \meter}$.
Since motor speed measurements are not available, thrust and torques feedback are estimated through a battery drain model \cite{Dhaybi2020_Quadrotor_Estimation}.
\subsection{Testing Scenario} \label{section_test_scenario}
\begin{figure}
\centerline{\includegraphics[width=3.34in]{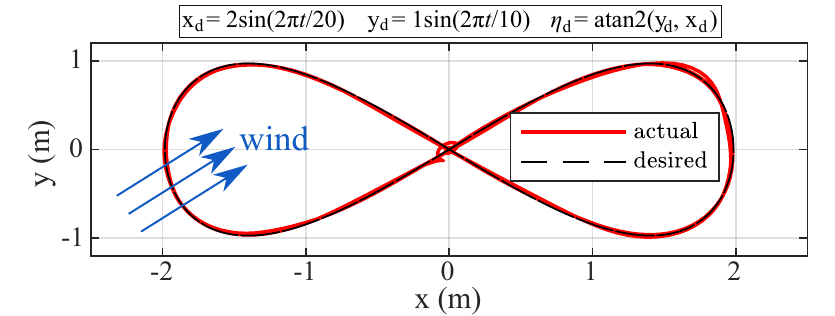}}
\caption{Tracking trajectory of the quadrotor in the simulation test after fine-tuning the ICS.}
\label{fig_XY_ICS_Sim}
\end{figure}

The controller is tested in the following scenario: the quadrotor, initially unladen, carries a payload  and takes-off from the ground, it performs three $\infty$-shape loops, then lands in its initial location. 
A payload with mass $m_{\mathrm{p}}=\SI[unitsep=medium]{0.2}{\kilogram}$ is placed beneath the original CoM of the quadrotor at coordinates $\bm{r}_{\mathrm{p}}=[r_{\mathrm{p}1},r_{\mathrm{p}2},r_{\mathrm{p}3}]=[50,50,-100]\,\SI[unitsep=medium]{}{mm}$ in $\mathcal{B}$.
The desired trajectory and the quadrotor's actual motion in the $\mathrm{x}$-$\mathrm{y}$ plane are shown in Fig. \ref{fig_XY_ICS_Sim}. 
The wind gust is applied at $t=\SI[unitsep=medium]{11}{s}$.
%
\subsection{Results}
\begin{figure}
\centerline{\includegraphics[width=3.34in]{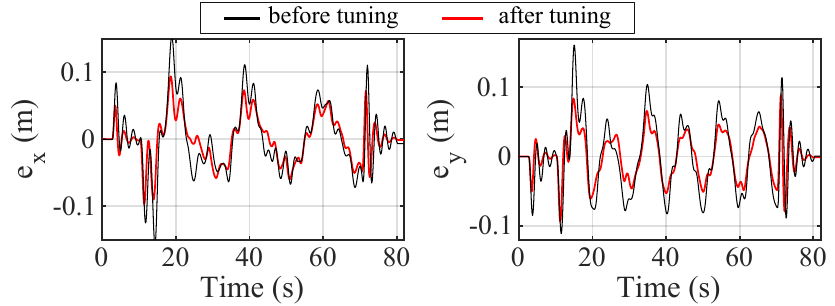}}
\caption{Lateral $\mathrm{x}$- and $\mathrm{y}$-motion tracking errors of the quadrotor in the simulation test before and after tuning the ICS.}
\label{fig_XY_error_improve}
\end{figure}

The controller gains are initially set as $\bm{k}_1=diag(1,1,2.6,30,30,4)$, $\bm{k}_2=diag(1,1,0.4,0.3,0.3,1)$, and $\bm{\gamma}=diag(0.4,0.4,0.4,1,1,1)$ to execute a preliminary performance check.
The overall performance of the ICS is deemed acceptable based on a mean absolute error ${MAE}\{[\mathrm{x},\mathrm{y},\mathrm{z}]\}$ of $[3.7,3.8,1.2]\,\SI[unitsep=medium]{}{\cm}$ in position and ${MAE}\{\eta_3\}=\SI[unitsep=medium]{1.5}{\degree}$ in heading.
However, the tracking error is still noticeable on the states $\mathrm{x}$ and $\mathrm{y}$ as shown in Fig.~\ref{fig_XY_error_improve}. 
\par
\begin{figure}
\centerline{\includegraphics[width=3.34in]{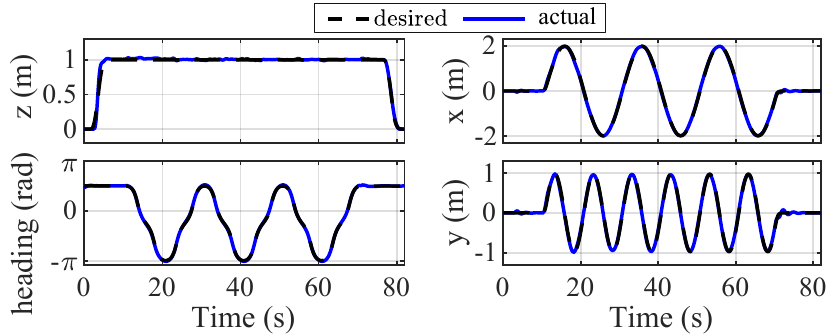}}
\caption{Tracking states of the quadrotor in the simulation test after fine-tuning the ICS.}
\label{fig_XYZ_yaw_ICS_Sim}
\end{figure}

By turning back to the method elaborated in Section~\ref{sec_PID_form}, the fine-tuning of the gains can be facilitated. 
First, we can calculate that the values of $k_{1,11}$ and $k_{2,11}$ refer to a combination of PD gains $k_{\mathrm{P}}=2.4$ and $k_{\mathrm{D}}=2$, then by referring to  (\ref{eq_kp_max}) and (\ref{eq_kD_min}), we see that $ k_{\mathrm{P},\mathrm{max}}=2.4$ and $ k_{\mathrm{D},\mathrm{min}}=2$.
By inspecting the performance in Fig.~\ref{fig_XY_error_improve}, we notice a considerable overshoot in the $\mathrm{x}$ and $\mathrm{y}$ motion.
Then, from PID tuning methods, we know that increasing $k_{\mathrm{D}}$ reduces overshoot, and most importantly, this does not violate the stability constraints derived from Lyapunov stability analysis as expressed in (\ref{eq_kD_min}).
Hence, by setting $k_{\mathrm{D}}=4$ and referring to (\ref{eq_K12_from_Kpd}), a new set of the adaptive gains are obtained as $k_{1,11}=3.7$ and $k_{2,11}=0.3$, which are utilized in the second simulation example. 
The final tracking performance is presented in Fig.~\ref{fig_XYZ_yaw_ICS_Sim}, and from Fig.~\ref{fig_XY_error_improve}, we see an improvement in the tracking performance of the fine-tuned ICS, 
where the mean absolute error reduces to $[2.6,2.6,1.1]\,\SI[unitsep=medium]{}{\cm}$. This results an error reduction of more than $\SI[unitsep=medium]{30}{\%}$ in the $\mathrm{x}$- and $\mathrm{y}$-motion due to the fine-tuned gains, as expected from the above analysis and without negatively influencing the system's stability, which demonstrates the proposed methodology's advantages. However, the reduction in the stability margins induced by lower $k_2$ values should be evaluated carefully.
\begin{remark} \label{rem_beyond_intuition}
    We note that the new combination of adaptive gains, $k_{1,11}$ and $k_{2,11}$, is not intuitive nor straightforward to guess from the initial set of gains and the corresponding system performance based on Lyapunov stability theory.
\end{remark}
\par
%
\subsection{Experimental Validation}
The proposed control system is implemented on a physical quadrotor, the Quanser QBall2 platform, and experimental validation is conducted in a testing facility equipped with a motion capture system, as shown in Fig.~\ref{fig_quadrotor_annotations}.
\par
\begin{figure}
\centerline{\includegraphics[width=3.34in]{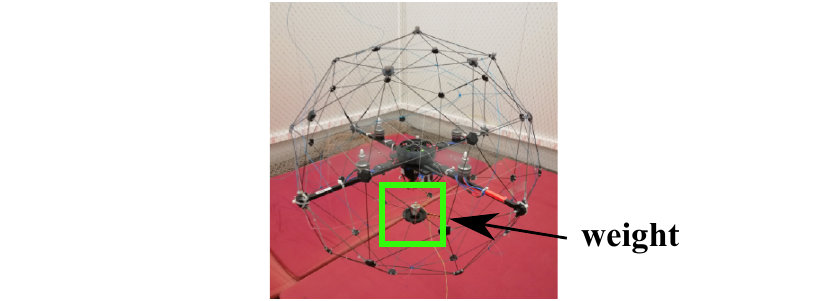}}
\caption{Demonstration of the Quanser QBall2 lifting a payload inside the UAV laboratory of the VRL at AUB.}
\label{fig_QBall2_in_Lab}
\end{figure}

The maneuver entails the quadrotor taking off vertically while carrying a payload of $m_{\mathrm{p}}=\SI[unitsep=medium]{0.2}{\kilogram}$, placed beneath the original CoM of the quadrotor at coordinates $\bm{r}_{\mathrm{p}}=[0,0,-100]~\SI[unitsep=medium]{}{mm}$ in $\mathcal{B}$, as demonstrated in Fig.~\ref{fig_QBall2_in_Lab}, then it moves sideways in the $\mathrm{x}$-$\mathrm{y}$ plane and drops the carried mass while moving, and finally performs landing.
\par
\begin{figure}
\centerline{\includegraphics[width=3.2in]{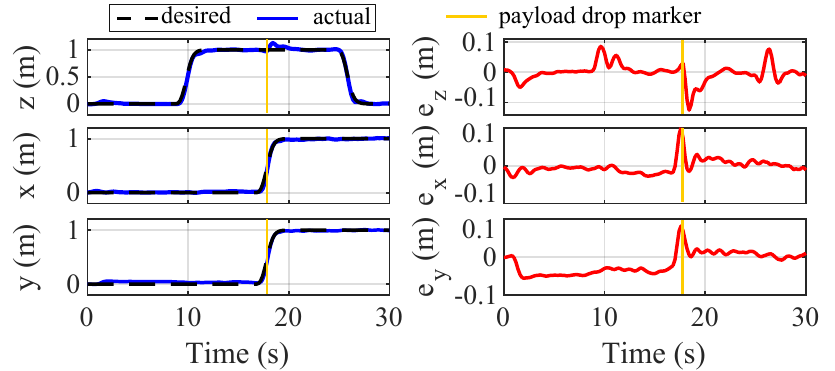}}
\caption{Tracking states (left) and errors (right) of the ICS-controlled quadrotor in the payload drop experiment.}
\label{fig_XYZ_exp}
\end{figure}

The tracking performance, shown in Fig.~\ref{fig_XYZ_exp}, is captured by the quadrotor's motion in the $\mathrm{z}$-, $\mathrm{x}$-, and $\mathrm{y}$-directions, respectively. The designed and fine-tuned ICS exhibits a stable and accurate tracking performance in following the desired trajectory, and is able to hold the quadrotor's position with minimal error in all directions, during lift-off from the ground and when the payload is dropped. 
The quadrotor promptly regains the desired height upon the sudden drop of the payload (at $t = \SI[unitsep=medium]{18}{s}$), and executes the remainder of the maneuver with accurate tracking performance.
%
%
\section{Conclusion}
\label{sec_conclusion}
We have presented a methodology for designing and tuning an adaptive and robust backstepping control law for a class of second-order nonlinear systems. The originally derived control law was reformulated as a two-DOF controller: a feedback component in a PID-like structure and a feedforward component that provides model compensation. The proposed formulation allows to gain insight, from a performance perspective, when dealing with backstepping-based control laws. In addition, we provide an attempt to answer a need for a simplified method to tune a relatively complex nonlinear control system, namely backstepping-based control, as compared to the ubiquitous PID controllers.
The obtained similarity draws a convenient link between the Lyapunov stability-based adaptive backstepping controller and a traditional PID controller. 
The proposed control and estimation scheme is validated in numerical simulation on a high-fidelity quadrotor model and experimentally on a real-life platform. 
Future work aims at extending the proposed method to other classes of nonlinear higher-order systems and other nonlinear control designs besides backstepping.


\section*{ACKNOWLEDGMENT}

This work is supported by the University Research Board (URB) at the American University of Beirut (AUB).

\bibliographystyle{ieeetr}

\bibliography{Ref/main}

\begin{thebibliography}{10}

\bibitem{Chang2002SelfTuningPID}
W.-D. Chang, R.-C. Hwang, and J.-G. Hsieh, ``A self-tuning {PID} control for a
  class of nonlinear systems based on the {L}yapunov approach,'' {\em Journal
  of Process Control}, vol.~12, no.~2, pp.~233 -- 242, 2002.

\bibitem{Astrom1995}
Astrom, {\em Adaptive Control}.
\newblock Addison-Wesley, New York, 1995.

\bibitem{Chang2009_PID_gain_selection}
P.~H. {Chang} and J.~H. {Jung}, ``A systematic method for gain selection of
  robust {PID} control for nonlinear plants of second-order controller
  canonical form,'' {\em IEEE Transactions on Control Systems Technology},
  vol.~17, pp.~473--483, March 2009.

\bibitem{Lee2014_PID_Backstepping_tuning}
J.~Y. {Lee}, M.~{Jin}, and P.~H. {Chang}, ``Variable {PID} gain tuning method
  using backstepping control with time-delay estimation and nonlinear
  damping,'' {\em IEEE Transactions on Industrial Electronics}, vol.~61,
  pp.~6975--6985, Dec 2014.

\bibitem{Han2009PIDtoADRC}
J.~{Han}, ``From {PID} to active disturbance rejection control,'' {\em IEEE
  Transactions on Industrial Electronics}, vol.~56, pp.~900--906, March 2009.

\bibitem{Yu2018BacksteppingNonlinear}
J.~Yu, P.~Shi, and L.~Zhao, ``Finite-time command filtered backstepping control
  for a class of nonlinear systems,'' {\em Automatica}, vol.~92, pp.~173 --
  180, 2018.

\bibitem{Yao1997}
B.~Yao and M.~Tomizuka, ``Adaptive robust control of {SISO} nonlinear systems
  in a semi-strict feedback form,'' {\em Automatica}, vol.~33, no.~5,
  pp.~893--900, 1997.

\bibitem{Yao2001MIMO}
B.~Yao and M.~Tomizuka, ``Adaptive robust control of {MIMO} nonlinear systems
  in semi-strict feedback forms,'' {\em Automatica}, vol.~37, pp.~1305--1321,
  2001.

\bibitem{Swaroop2000}
D.~{Swaroop}, J.~K. {Hedrick}, P.~P. {Yip}, and J.~C. {Gerdes}, ``Dynamic
  surface control for a class of nonlinear systems,'' {\em IEEE Transactions on
  Automatic Control}, vol.~45, pp.~1893--1899, Oct 2000.

\bibitem{Dong2012CommandFilteredBackstepping}
W.~{Dong}, J.~A. {Farrell}, M.~M. {Polycarpou}, V.~{Djapic}, and M.~{Sharma},
  ``Command filtered adaptive backstepping,'' {\em IEEE Transactions on Control
  Systems Technology}, vol.~20, pp.~566--580, May 2012.

\bibitem{Wen2011RAC_nonlinear}
C.~{Wen}, J.~{Zhou}, Z.~{Liu}, and H.~{Su}, ``Robust adaptive control of
  uncertain nonlinear systems in the presence of input saturation and external
  disturbance,'' {\em IEEE Transactions on Automatic Control}, vol.~56,
  pp.~1672--1678, July 2011.

\bibitem{Benaskeur2002_BacksteppingBasedPID}
A.~Benaskeur and A.~Desbiens, ``Backstepping-based adaptive {PID} control,''
  {\em IEE Proceedings - Control Theory and Applications}, vol.~149,
  pp.~54--59(5), January 2002.

\bibitem{Skjetne2004_IntegralControlBackstepping}
R.~{Skjetne} and T.~I. {Fossen}, ``On integral control in backstepping:
  analysis of different techniques,'' in {\em Proceedings of the 2004 American
  Control Conference}, vol.~2, pp.~1899--1904, June 2004.

\bibitem{Mian2008_BacksteppingBasedPID}
A.~A. {Mian}, M.~I. {Ahmad}, and D.~{Wang}, ``Backstepping based {PID} control
  strategy for an underactuated aerial robot,'' {\em IFAC Proceedings Volumes},
  vol.~41, no.~2, pp.~15636 -- 15641, 2008.

\bibitem{Kartal2019Backstepping_PID_UAV}
Y.~{Kartal}, P.~{Kolaric}, V.~{Lopez}, A.~{Dogan}, and F.~{Lewis},
  ``Backstepping approach for design of {PID} controller with guaranteed
  performance for micro-air {UAV},'' {\em Control Theory and Technology},
  vol.~18, pp.~19--33, Feb. 2019.

\bibitem{Pan2010Robust2ndOrderLyapunov}
Y.~{Pan}, G.~{Liu}, and K.~D. {Kumar}, ``Robust stability analysis of
  asymptotic second-order sliding mode control system using {L}yapunov
  function,'' in {\em The 2010 IEEE International Conference on Information and
  Automation}, pp.~313--318, June 2010.

\bibitem{Hu2010DIARC_Gantry}
C.~Hu, B.~Yao, and Q.~Wang, ``Integrated direct/indirect adaptive robust
  contouring control of a biaxial gantry with accurate parameter estimations,''
  {\em Automatica}, vol.~46, no.~4, pp.~701 -- 707, 2010.

\bibitem{Yao2002DIARC}
B.~Yao and R.~Y. Dontha, ``Integrated direct/indirect adaptive robust precision
  control of linear motor drive system with accurate parameter estimations,''
  {\em IFAC Mechtronics Systems}, vol.~35, no.~2, pp.~587--592, 2002.

\bibitem{Kourani2018}
A.~Kourani, K.~Kassem, and N.~Daher, ``Coping with quadcopter payload variation
  via adaptive robust control,'' in {\em 2018 IEEE Int. Multidisciplinary Conf.
  on Engineering Technology}, 2018.

\bibitem{Zhang2019}
J.~Zhang, D.~Gu, Z.~Ren, and B.~Wen, ``Robust trajectory tracking controller
  for quadrotor helicopter based on a novel composite control scheme,'' {\em
  Aerospace Science and Technology}, vol.~85, pp.~199--215, 2019.

\bibitem{Dhaybi2020_Quadrotor_Estimation}
M.~{Dhaybi} and N.~{Daher}, ``Accurate real-time estimation of the inertia
  tensor of package delivery quadrotors,'' in {\em 2020 American Control
  Conference (ACC)}, pp.~1520--1525, July 2020.

\end{thebibliography}

\end{document}